\documentclass[twocolumn,showpacs,preprintnumbers,amsmath,amssymb]{revtex4}

\usepackage{graphicx}
\usepackage{bm}

\begin{document}

\title{Complete population transfer in degenerate $n$-state atoms}

\author{J.H. McGuire}
\author{Kh.Kh. Shakov}
\author{Kh.Yu. Rakhimov\footnote{Permanent address: 
Department of Heat Physics, Uzbekistan Academy of Sciences, 
28 Katartal St., Tashkent 700135, Uzbekistan}}
\affiliation{Physics Department, Tulane University, New Orleans, LA 70118 USA.}

\date{\today}

\begin{abstract}
We find a set of conditions to achieve complete population transfer,
via coherent population trapping, from an initial state to a designated
final state at a designated time in a degenerate $n$-state atom, where
the transitions are caused by an external interaction.
\end{abstract}


\maketitle


Population control in quantum systems, namely transfer of electrons
from an ensemble of atoms all in the same initial state to specified
final states within the ensemble, is used in problems ranging from coherent 
population trapping \cite{alz76,metcalf02}, 
including electromagnetically induced transparency \cite{harris97,lukin01}, 
and quantum computing 
\cite{Nielsen,Bouwmeester},   
to chemical dynamics \cite{Rabitz93,Rabitz03}.    
These problems are modeled in terms of an $n$-state atom 
interacting with an external field 
\cite{Shore,Milonni88,Meystre}. 
In recent papers \cite{sm02,msr03,rms03} we have shown that 
2-state and 3-state atoms are relatively easy to understand 
in the degenerate limit, where all states have the same energy.
We have specifically shown how to achieve complete population transfer in  
degenerate 2-state \cite{sm02} and 3-state \cite{rms03} atoms 
when the matrix elements of the external interaction have a common time dependence.
Here we present a way to achieve complete population transfer in degenerate 
$n$-state atoms. 

Our degenerate energy approximation is somewhat similar to the
rotating wave approximation (RWA) \cite{Shore,Milonni88} that
has been widely applied to both $2$ and $3$-state atomic models.
In RWA, however, degenerate atomic states are not used.
Instead one tunes the frequency of the external field, $V(t)= \chi \cos(\omega t)$,
to the frequency difference of two non-degenerate levels
so that the detuning parameter, $\Delta = \hbar \omega + E_1 - E_2$,
tends to zero.  Thus in RWA an initial state of an atom plus one
photon is degenerate in energy with the final state of the atom.
An advantage of using degenerate atomic states is that one is not restricted
to external interactions with frequencies close to the transition frequency.
Thus $\omega$ can be used for control \cite{sm02}, e.g. to vary the duration of time
that the transferred population remains in the designated state, or to reduce
the population leakage that occurs when the energy levels are not fully degenerate.
On the other hand RWA has the advantage that analytic solutions
have been found when the detuning parameter is small, but finite.


Consider an $n$-state atom interacting with an external field, $V_{ext}(\vec{r},t)$.
The total Hamiltonian for this system is $H = H_0 + V_{ext}(\vec{r},t)$.  The $n$ eigenstates, 
$\phi_k$, and corresponding eigenenergies, $E_k$, of $H_0$ are assumed to be known.
The total wavefunction may be expanded in terms of the known eigenstates, namely, 
$\Psi(t) = a_1(t) \phi_1 + a_2(t) \phi_2 + \cdots + a_n(t) \phi_n$.
With atomic units, using $i \dot{\Psi} = (H_0 + V_{ext}(\vec{r},t)) \Psi$, 
with $H_0 \phi_k  = E_k \phi_k$ and $\int \phi^*_k \phi_j d\vec{r} = \delta_{kj}$,
one then obtains  \cite{Milonni88},   
$i \dot{a}_k(t) = E_k a_k(t) + \sum_{j=1}^n V_{kj}(t) a_j(t)$,  
where $V_{kj}(t) = \int \phi^*_k V_{ext}(\vec{r},t) \phi_j d \vec{r}$. 
These equations are exact for an $n$-state atom.
We now assume that the system is degenerate, namely that
all the energies, $E_k$, are the same. 
Since the zero point of energy is arbitrary, one may generally set $E_k = 0$.
These conditions give the coupled equations for a degenerate $n$-state system, namely,
\begin{eqnarray}
\label{ndegen}
i \dot{a}_k(t) =  \sum_{j =1}^n V_{kj}(t) a_j(t)  \ \ .
\end{eqnarray}
We use the initial condition $a_1(0) = 1$, and $a_k(0) = 0$ for $k \neq 1$.
We additionally assume that all of the $V_{kj}(t)$ have the same time dependence.
Here we also take $V_{kj}(t) = V_{jk}(t)$ to be real.


As will be verified below, there is a general scheme 
to find analytic solutions for degenerate $n$-state atoms \cite{msr03}.
This scheme is straightforward.
Seek a solution to Eqs(\ref{ndegen}) of the form
$c(t) = x_1 a_1(t) + x_2 a_2(t) + \cdots + x_n a_n(t)$. 
Since $a_i(0) = \delta_{i1}$, one has that $x_1 = 1$.  Next calculate
$i \dot{c}(t)$ using Eqs(\ref{ndegen}) and require that $i \dot{c}(t) = z V(t) \ c(t)$.
Here $V(t)$ is a common factor for the $V_{kj}(t)$ terms,
and $z$ is a linear sum of the $x_i$'s, dependent
on the relative (time independent) strengths of the $V_{kj}(t)$.
As illustrated below, this leads to an $n^{th}$ order equation in each of 
the $x_i$'s for $i \neq 1$, whose roots may be denoted by $x_{ji}$.
This yields $n$ eigenvalues, $z_j$, and $n$ eigenfunctions,
$c_j(t) = e^{- i z_j A(t)}$, where $A(t) = \int_0^t V(t')dt'$.
This process determines the matrix elements, ${\cal M}_{jk}$,
for $c_j(t) = \sum_k^n {\cal M}_{jk} a_k(t)$.  Specifically, ${\cal M}_{jk} = x_{jk}$.
Inverting this relation yields the probability amplitudes for the
electron population, $a_k(t) = \sum_j^n {\cal M}^{-1}_{kj} c_j(t)
= \sum_j^n {\cal M}^{-1}_{kj} e^{-i z_j A(t)}$.
Using $\cos(a-b) = \cos a \cos b + \sin a \sin b$,  one quickly obtains,
\begin{eqnarray}
\label{Pkn}
P_k(t) = |a_k(t)|^2 = \sum_{i}^n \ \sum_{j}^n  \
{\cal M}_{ki}^{-1}{\cal M}_{kj}^{-1}  \cos[(z_i - z_j) A(t)] .
\end{eqnarray}
Thus the occupation probabilities may be determined analytically for a degenerate $n$-state atom.

Since the $x_{ji}$'s and $z_j$'s vary with the $V_{kj}(t)$,
one may seek conditions on the matrix elements $V_{kj}(t)$
and on $A(t_0)$ such that the electron populations $P_k(t_0) = |a_k(t_0)|^2$
take desired values at $t = t_0$.
It has been shown that complete population transfer
occurs at $t = t_0$ if $A(t_0)/\pi = 1/2$ in the 2-state atom \cite{sm02}
and $A(t_0)/\pi = 1 / \sqrt{2}$ in the 3-state atom \cite{rms03}.  
In addition $V_{13}(t)/V_{23}(t) = 1$, and all the $V_{ii}$ are the same for the 3-state atom.   


Since for $k \neq 1$ all the $a_k(0) = 0$, the degenerate $n$-state equations, Eqs(\ref{ndegen}), 
simplify if $V_{kj}(t) = \gamma V_{23}(t)$ for all $j \geq 3$.  
Using $V(t) = \alpha^{-1} V_{12} 
= \beta^{-1} V_{13} = \gamma^{-1} V_{23} = \epsilon_i^{-1} V_{ii}$, 
Eqs(\ref{ndegen}) become,
\begin{eqnarray}
\label{nsym}
i \dot{a}_1(t) &=& V(t) \ ( \ \epsilon_1 \ a_1(t) + \alpha \ a_2(t) + (n-2) \beta \ a_3(t) \ )  
	  , \\ \nonumber
i \dot{a}_2(t) &=& V(t) \ ( \ \alpha \ a_1(t) + \epsilon_2 \ a_2(t) + (n-2) \gamma \ a_3(t) \ )
	\  ,  \\ \nonumber
i \dot{a}_3(t) &=& V(t) \ ( \ \beta \ a_1(t) + \gamma \ a_2(t) + ( \epsilon_3 + (n-3) ) \ a_3 \ ) 
	\  .
\end{eqnarray}
This partially symmetric $n$-state system is mathematically equivalent to
a 3-state system with $\tilde{V}_{13} = (n-2)\tilde{V}_{31}$ and
$\tilde{V}_{23} = (n-2) \tilde{V}_{32}$.  
Since $V(t)$ is arbitrary at this point, we may set $\gamma = 1$ without loss of generality.

These equations may be solved using the general scheme described above.  
To find a solution that is mathematically simple, following the
3-state atom \cite{rms03} we choose $\beta = 1$ and all $\epsilon_j = \epsilon$.  
The value of $\epsilon$ may be arbitrarily changed by
an overall phase transformation of the $a_j$.  
Here we use $\epsilon = 0$.  
Taking $x_2 = x$ and $x_3 = y$ to simplify notation, one quickly obtains,
$x = (\alpha + y)/(\alpha x + (n-2)y)$ and 
$y = ( 1 + x + (n-3) y )/(\alpha x + (n-2)y)$ with $z = \alpha x + (n-2) y$. 
This yields cubic equations in $x$ and $y$.
However, it is evident in this case that if $x = -1$ then $y = 0$, and
if $x = 1$ then $y = y_{\pm} = \frac{1}{2}(\frac{-\alpha + n - 3}{n-2} 
\pm \sqrt{(\frac{\alpha - n + 3}{n-2})^2 + \frac{8}{n-2}} \ )$.  
Hence there are three eigenvalues for $x$, $y$ and $z$, namely $\{x_j\} = \{1,1,-1\}$,
$\{y_j\} = \{y_+,y_-,0\}$ and $\{z_j\} = \{ \alpha + (n-2) y_+, \alpha + (n-2)y_-, -\alpha\}$.
This gives three eigenfunctions, $c_j = e^{-iz_j A(t)}$, 
which are linear combinations of the $a_k(t)$.
Specifically $a_k = \sum_{j = 1}^3 {\cal M}^{-1}_{kj} c_j$, where,
$ {\cal M}^{-1} =\frac{1}{2(y_+ - y_-)}\left(\begin{array}{ccc}
      -y_-      & y_+              &   (y_+ - y_-)  \\
      -y_-      & y_+              &  -(y_+ - y_-)  \\
 \frac{2}{n-2}  &  \frac{-2}{n-2}  &      0         \\
\end{array}\right) $.
This yields transition probabilities $P_k(t) = |a_k(t)|^2$, in accord with Eqs(\ref{Pkn}). 


From Eqs(\ref{Pkn}) extrema of $P_i(t)$ occur at $t = t_0$ when
$(z_1 - z_2) A(t_0)/\pi = k$ and $(z_2 - z_3) A(t_0)/\pi = k'$.
The third condition is redundant since $P_1 + P_2 + (n-2)P_3 = 1$. 
One may show after some algebra that these conditions are met
when,
\begin{eqnarray}
\label{qcond}
  A(t_0)/\pi &=& \pm n_0 \sqrt{\frac{9}{18 (n-2) + 4 (n-3)^2}} \ \ , \nonumber \\ 
 \alpha &=& V_{12}(t)/V_{23}(t) =  -\frac{1}{3} (n-3)  \ \ , \nonumber \\ 
 \beta  &=& V_{13}(t)/V_{23}(t) = 1  \ \ ,
\end{eqnarray}
where $n$ is the number of degenerate states, and
$n_0$ is any odd integer. It may be shown that $k = -2k' = 2n_0$.
Using Eqs(\ref{Pkn}) one can then show, 
\begin{eqnarray}
\label{Pn}
  P_1 &=& |a_1(t)|^2 = \frac{1}{8} [ 3 + \cos(\theta) + 4 \cos(\theta/2) ]  \ \ , \nonumber \\
  P_2 &=& |a_2(t)|^2 = \frac{1}{8} [ 3 + \cos(\theta) - 4 \cos(\theta/2) ]  \ \ , \nonumber \\
  P_3 &=& |a_3(t)|^2 = (\frac{1}{n-2})\frac{1}{2} \sin^2(\theta/2) \ \ ,
\end{eqnarray}
where $\theta = \theta(t) = 2 \pi n_0 [A(t)/A(t_0)]$  with $A(t) = \int_0^t V(t') dt'$.
Complete population transfer occurs from state 1 to state 2 at $t = t_0$
when $\theta(t) = 2 \pi n_0$.
Using Eqs(\ref{qcond}) one may confirm that as $n \to \infty$ these solutions reduce to 
those of a 2-state atom with $V_{11} = V_{22}$, where $P_1 = \cos^2A(t)$ and $P_2 = \sin^2A(t)$.


To verify both the general scheme and our algebra, we solved Eqs(\ref{ndegen})
numerically for a 4-state atom using a fourth order Runge Kutta method for
$V(t) = \chi \cos(\omega t)$.
The results, shown in the figure, are identical to those given analytically by Eqs(\ref{Pn}).
When the numerical code is run with non-degenerate energies, 
i.e. $E_i - E_j = \hbar \omega_{ij} \neq 0$, we find that 
population transfer is not complete.  When $(\omega_{ij}/\omega)^2 < 1$
the population leakage varies as $c(\omega_{ij}/\omega)^2$, where $|c| < 1$. 
Thus, population leakage may be reduced by increasing the frequency, $\omega$,
of the external field.

\begin{figure}
\scalebox{0.7}{\includegraphics{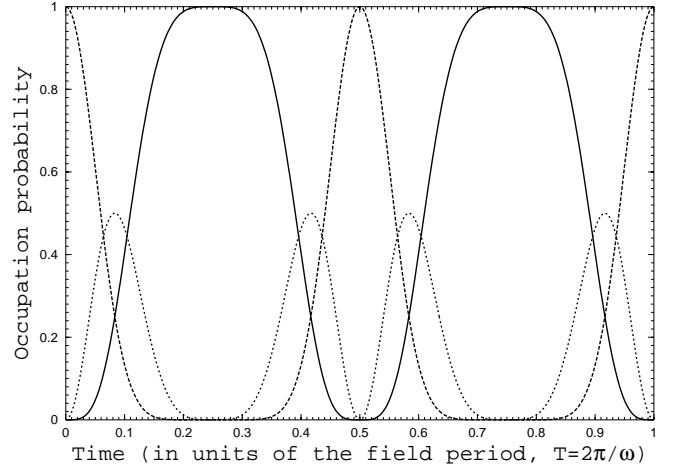}}
\caption{Occupation probabilities as a function of time in a degenerate $n$-state atom
in an external field $V(t) = \chi \cos(\omega t)$.
The long dash line denotes $P_1(t)$, the probability that the electron is in its initial state;
the solid line denotes $P_2(t)$, the probability that the electron is in the target state; 
and the short dash line denotes $(n-2) P_3(t)$, where $P_3(t)$ is the probability that 
the electron is a state other than the initial or target state.  
Complete population transfer occurs at $t = t_0$ when 
in Eqs(\ref{Pn}), $\theta = 2 \pi n_0 A(t)/A(t_0)$.
} 
\label{poptfig1}
\end{figure}


An advantage of our approach is that one is free to change the
shape of $V(t)$.  This may be used to control how long the system remains
in its target state.  An extreme example is a sudden `kick', where
$V(t) = A_0 \delta(t - t_0)$. In this case the system changes suddenly
from state 1 to state 2 at $t = t_0$ and remains there indefinitely.
This is a switch.  One may apply a series of switches by taking
$V(t) = A_0 [ \ \delta(t - t_1) + \delta(t - t_2) + .... + \delta(t - t_k) \ ]$.
Alternatively one may switch to different states by adjusting the
matrix elements so that after each step $V_{12}(t)$ is interchanged with $V_{1'2}(t)$,
where state $1'$ is now the state occupied before the next `kick' is applied.


In this paper we consider transitions amplitudes and
probabilities in an $n$-state atom with degenerate states
coupled by $V_{ext}(t)$ whose matrix elements, $V_{ij}(t)$
all have a common time dependence.  
If the states are truly degenerate, the transition probabilities
are given exactly by Eqs(\ref{Pkn}) above.  Moreover, under conditions detailed
in this paper population transfer can be complete.  RWA differs in these two
features since in RWA correction terms are always present for any finite value of $\omega$.
On the other hand, in realistic atomic systems while the $V_{ij}(t)$ often all have a common 
time dependence, the energy levels are seldom, if ever, exactly degenerate.
Also levels outside $n$-state manifold usually exist.
Consequently, use of our results is restricted to external fields
with frequencies in the range, $\omega_{min} < \omega <  \omega_{max}$.
Here $\hbar \omega_{min}$ is the energy splitting of the nearly
degenerate states, and $\hbar \omega_{max}$ is the energy difference
between the nearly degenerate $n$-state manifold and the closest state in energy outside
the manifold.  While this restriction prevents the use of ideal 'kicks', 
if $\omega_{max}$ is large enough practical 'kicks' with a finite
width in time may be designed that approximately satisfy all necessary requirements.

In general complete population transfer can be achieved by our scheme
if all but one of the transition matrix elements are the same.     
This means that interactions obeying dipole selection rules 
may be used only in 2-state systems, and in 3-state degenerate systems 
where, fortuitously, $\alpha = n-3 = 0$. 
We remark that use of degeneracy, $\Delta E \to 0$,
imposes $\Delta t \to \infty$.  This removes time sequencing in
intermediate steps of the reaction process \cite{rms03}.
That is, 
any physical effect due to time sequencing of intermediate interactions 
is lost in the limit of degeneracy.

In summary,  in degenerate $n$-state atoms electron population
is completely transferred via an external interaction at a designated time, $t_0$,
from a launch state (1) to a target state (2) under two conditions.  
The first condition for complete transfer is that the ratio of the matrix elements 
of the external interaction, $V_{kj}(t)$, all be the same except for 
$V_{12}(t)/V_{23}(t) = \alpha = -\frac{1}{3} (n-3)$.
The second condition is that at $t = t_0$ the phase area of $V(t)$ satisfy
$A(t_0) = \int_0^{t_0} V(t') dt' = 
\pm n_0 \pi \  \sqrt{\frac{9}{18 (n-2) + 4 (n-3)^2}}$,
where $V_{ij}(t) = V(t)$ is any one of the matrix elements except $V_{12}(t)$.
The ensuing equations for the populations of the states are simple.
We have found other, more complex, solutions to Eqs(\ref{ndegen}) 
that yield complete population transfer. 
The duration of time the transferred population remains in state 2
can be controlled by varying the shape of $V(t)$.

\begin{acknowledgments}
We thank J.H. Eberly and B.W. Shore for useful discussion.
This work was supported in part by the Division of Chemical Sciences, Office
of Sciences, U.S. Department of Energy.  KhR is supported by a NSF-NATO Fellowship.
\end{acknowledgments}


\end{document}